\documentclass[aps,superscriptaddress,preprint]{revtex4}

\usepackage{titlesec}
\usepackage{amssymb,amsmath,amsfonts,latexsym,graphicx,epsfig}
\usepackage{empheq}
\usepackage{epstopdf}
\usepackage{float}
\usepackage{xcolor}

\newcommand{\bea}{\begin{eqnarray}}
\newcommand{\ena}{\end{eqnarray}}
\newcommand{\be}{\begin{equation}}
\newcommand{\en}{\end{equation}}
\newcommand{\nn}{\nonumber\\}
\newcommand{\ed}{\end{document}}

\newcommand{\la}{\langle}
\newcommand{\ra}{\rangle}
\newcommand{\Bla}{\Big\langle}
\newcommand{\Bra}{\Big\rangle}
\newcommand{\Tr}{\mbox{\rm{tr}}}
\newcommand{\Heff}{\mbox{${\cal H}_{\rm eff}$}}

\begin{document}

\title{Weak nonleptonic decays of vector B-mesons}

\author{Mikhail~A.~Ivanov}
\affiliation{Bogoliubov Laboratory of Theoretical Physics,
	Joint Institute for Nuclear Research, Dubna, Russia}

\author{Zhomart~Tyulemissov}
\affiliation{Bogoliubov Laboratory of Theoretical Physics,
	Joint Institute for Nuclear Research, Dubna, Russia}
\affiliation{ The Institute of Nuclear Physics, Ministry of Energy of
	the Republic of Kazakhstan, 050032 Almaty, Kazakhstan} 
\affiliation{ Al-Farabi Kazakh National University, 050040 Almaty, Kazakhstan}

\author{Akmaral~Tyulemissova}
\affiliation{Bogoliubov Laboratory of Theoretical Physics,
	Joint Institute for Nuclear Research, Dubna, Russia}
\affiliation{ The Institute of Nuclear Physics, Ministry of Energy of
	the Republic of Kazakhstan, 050032 Almaty, Kazakhstan} 
\affiliation{ Al-Farabi Kazakh National University, 050040  Almaty, Kazakhstan}

\begin{abstract}
We study the radiative and weak nonleptonic decays of vector B-mesons
within the covariant confined quark model (CCQM) developed in our previous
papers. First, we calculate the matrix elements and decays widths
of the radiative decays $B^\ast\to B\gamma$. The obtained results are compared
with those obtained in other approaches. Then we consider the
nonleptonic decays $B^\ast\to D^\ast V$ which proceed via tree-level quark
diagrams. It is shown that the analytical expressions for the amplitudes
correspond to the factorization approach. In the framework of our model
we calculate the leptonic decay constants and the form factors of
the $B^\ast\to D^\ast(V)$ transitions in the entire physical region of the
momentum transfer squared. Finally, we calculate the two-body decay widths
and compare our results with other models. 
\end{abstract}


\maketitle

\section{Introduction}
\label{sec:intro}



In 2009, the Belle~\cite{Belle:2008ezn} collaboration has reported
on the determination of the masses of the $B_s$ and $B_s^\ast$ mesons
$m_{B_s}= 5364.4\pm 1.5$~MeV and $m_{B^\ast_s}= 5416.4\pm 0.6$~MeV.
The LHCb~\cite{LHCb:2012iuq} collaboration announced the discovery of
a vector $B^\ast$-meson with the mass $m(B^\ast) = 5324.26(41)$~MeV.

Since the $ b (c,s,d,u) $ mesons cannot annihilate into gluons,
the excited states decay to the ground state via the cascade emission of
photons or pion pairs, leading to total widths that are less than
a few hundred keV. The difference between masses
$m_{B^\ast} - m_{B} = 45.21 \pm 0.21 $~MeV, a
$ m_{B^\ast_s} - m_{B_s}  = 48.5^{+1.8}_{-1.1}$~MeV less than the mass of
lightest meson ($\pi$-meson), ($m_\pi = 134.98$ MeV), as a result of which
these mesons cannot decay through a strong channel. Consequently, for vector
$ B $-mesons, radiative decays will be dominant. As is known, weak
nonleptonic decays of $B^\ast$-mesons are always suppressed in comparison
with their electromagnetic decay~\cite{LHCb:2012iuq}. However, due to
small cross sections, these decays have not yet been detected experimentally.
The situation can be improved with the help of the LHC and
Belle-II experiments, since the annual integrated luminosity of Belle-II
is expected to reach $\approx 13$ ab$^{-1}$ and this makes it possible to
detect weak decays with branchings greater than $\mathcal{O}(10^{-9})$.
Moreover, the LHC experiment will also provide new experimental data for weak
decays of $B^\ast$-mesons, due to the large production cross section for
$b$-quarks.


A search for excited states of the $B_c^\pm$-meson
has been started at 2014 when the ATLAS collaboration reported
on the observation of a new state~\cite{ATLAS:2014lga} through
its hadronic transition to the ground state $B_c^\pm$.
The mass of the observed state was found to be $ 6842(6)$~MeV.
The mass and decay of this state are consistent
with expectations for the second S-wave state of the  ground $B_c$~state
denoted as $B_c(2^1S_0)$ or $B_c(2S)$.
%
%
Search for excited $B_{c}^{+}$ states was started at LHCb~\cite{LHCb:2017rqe}.
In 2019, the CMS collaboration observed two states $B_c(2S)$ and
$B^\ast_c(2S)$~\cite{CMS:2019uhm} in the $B_c\pi^+\pi^-$ invariant mass spectrum.
They are separated in mass by $\approx 29$~MeV. The mass of the $B_c(2S)$
is measured to be $6871.0 \pm 1.6$~MeV.
%
%
The LHCb collaboration has reported on  the observation of an excited
$B_c$ state in the the $B_c\pi^+\pi^-$ invariant mass
spectrum~\cite{LHCb:2019bem}. The observed peak has a mass of
$6841.2 \pm 1.0$~MeV. It is consistent with expectations of
the $B_c^\ast(2^3S_1)$ state. A second state is seen with a mass of
$6872.1 \pm 1.5$~MeV, and is consistent with the $B_c(2^1S_0)$ state.



Radiative and hadronic heavy meson decays
of the heavy vector $B^\ast$-mesons have been  evaluated
using the Heavy Quark Effective Theory and the Vector Meson Dominance
hypothesis~\cite{Colangelo:1993zq}. It was found that
$\Gamma(B^{\ast\,+}\to B^+\gamma)     = 220(90)$~eV and
$\Gamma(B^{\ast\,0}\to B^0\gamma)     =  75(27)$~eV.
%
%
The radiative and hadronic decays of vector heavy mesons were analyzed
within the relativistic quark model with confined light
quarks~\cite{Ivanov:1994ji}. The following results were obtained
for the value of the constituent bottom quark mass $m_b=5$~GeV,
$\Gamma(B^{\ast\,+}\to B^+\gamma)     = 401$~eV and
$\Gamma(B^{\ast\,0}\to B^0\gamma)     = 131$~eV. 
%
%
The method of QCD sum rules in the presence of the external electromagnetic
field was used to analyze radiative decays of charmed or bottomed mesons
\cite{Zhu:1996qy}. The calculated values of the decay widths were found to be
$\Gamma(B^{\ast\,+}\to B^+\gamma)     = 380(60) $~eV,
$\Gamma(B^{\ast\,0}\to B^0\gamma)     = 130(30) $~eV and
$\Gamma(B_s^{\ast\,0}\to B_s^0\gamma) =  220(40) $~eV. 
%
%
Radiative transitions in heavy mesons  have been considered in a
relativistic quark model~\cite{Goity:2000dk}.
The calculated values of the decay widths with the model parameter
$\kappa^q=0.55$) were found to be
$\Gamma(B^{\ast\,+}\to B^+\gamma)     = 740(88)$~eV,
$\Gamma(B^{\ast\,0}\to B^0\gamma)     = 228(27)$~eV,
$\Gamma(B_s^{\ast\,0}\to B_s^0\gamma) = 136(12)$~eV. 
%
%
Radiative magnetic dipole decays of heavy-light vector mesons into
pseudoscalar mesons have been considered within
the relativistic quark model~\cite{Ebert:2002xz}.
The results  for the mixture of vector and scalar confining potentials
with the mixing parameter $\varepsilon=-1$ look as  
$\Gamma(B^{\ast\,+}\to B^+\gamma)     = 190$~eV,
$\Gamma(B^{\ast\,0}\to B^0\gamma)     =  70$~eV,
$\Gamma(B_s^{\ast\,0}\to B_s^0\gamma) =   54$~eV.
%
%
Decay constants and radiative decays of heavy mesons in light-front quark
model~\cite{Choi:2007se} were found to be
$\Gamma(B^{\ast\,+}\to B^+\gamma)     = 400(30)$~eV,
$\Gamma(B^{\ast\,0}\to B^0\gamma)     = 130(10) $~eV,
$\Gamma(B_s^{\ast\,0}\to B_s^0\gamma) =  68(17)  $~eV. 
%
%
In this paper~\cite{Chang:2019xtj},  the light-front quark model (LFQM)
was employed to evaluate the decay widths:
$\Gamma(B^{\ast\,+}\to B^+\gamma)     = 349(18)$~eV,
$\Gamma(B^{\ast\,0}\to B^0\gamma)     =  116(6)$~eV,
$\Gamma(B_s^{\ast\,0}\to B_s^0\gamma)  =   84(10)$~eV.
%
%
%
The vector $B^\ast_c$ meson physics has attracted theorists from all over
the world since 1994. First attempt to study $B_c^\ast$ meson
was made in the framework of nonrelativistic quarkonium quantum
mechanics by using the QCD-motivated potential~\cite{Eichten:1994gt}.
It was found that $\Gamma(B_c^{\ast\,+}\to B_c^+\gamma) =   135$~eV.
Later on the authors updated this study and published the new
work~\cite{Eichten:2019gig}.
%
%
In the framework of potential models for heavy quarkonium the mass
spectrum for the system $(\bar b c$) was considered~\cite{Gershtein:1994dxw}.
Spin-dependent splittings, taking into account a change of a constant
for effective Coulomb interaction between the quarks, and widths of
radiative transitions between the  $(\bar b c$) levels were
calculated $\Gamma(B_c^{\ast\,+}\to B_c^+\gamma) =   60$~eV.
%
%
In the paper~ \cite{Ebert:2002xz}, the decay width was found to be
$\Gamma(B_c^{\ast\,+}\to B_c^+\gamma) =   33$~eV. 
%
%
The homogeneous bag model was employed in Ref.~\cite{Liu:2022bdq}
to calculate the width of the radiative decay
$\Gamma(B_c^{\ast\,+}\to B_c^+\gamma) = 53(3)$~eV.
The spectrum of heavy mesons including the excited states
was treated in the framework of the heavy quark effective theory
in Ref.~~\cite{Zeng:1994vj}.
The $B_c$ spectroscopy was investigated in a quantum-chromodynamic
potential model by using a quantum-chromodynamic potential
model~\cite{Gupta:1995ps}, within the  lattice Non-Relativistic QCD
(NRQCD)~\cite{Davies:1996gi},
Phenomenological predictions of the properties of the $B_c$ system
have been done in Ref.~~\cite{Fulcher:1998ka} by using Richardson’s potential.
In particular, the width of the radiative decay of $B_c^\ast$
was found to be $\Gamma(B_c^{\ast\,+}\to B_c^+\gamma) = 59$~eV.
The properties of heavy quarkonia and $B_c$ mesons have been studied
in the relativistic quark model~\cite{Ebert:2002pp}.
In Ref.~~\cite{Godfrey:2004ya} the spectrum of the charm-beauty mesons
and their radiative decays were studied  by using the relativized quark
model~\cite{Godfrey:2004ya}. It was found that
$\Gamma(B_c^{\ast\,+}\to B_c^+\gamma) = 80$~eV.
The observation possibility of $B_c$~excitations at LHC was discussed
in Ref.~\cite{Berezhnoy:2013sla}.

So far no experimental measurement of the vector $ B^\ast_c$-meson mass
available. However there are a few reliable calculations made
in the Lattice QCD. In the paper~\cite{Gregory:2009hq}
the prediction was done to be $m(B^*_c) = 6330(7)(2)(6)$~MeV
by using the Highly Improved Staggered Quark formalism
to handle charm, strange and light valence quarks in full lattice QCD, and
NRQCD.
The  improved results for the $B$ and $D$ meson spectrum
from lattice QCD including the effect of $u/d$, $s$ and $c$ quarks
in the sea were presented in Ref.~\cite{Dowdall:2012ab}. It was found that
$m(B^*_c) = 6278(9)$~MeV.
Precise predictions of charmed-bottom hadrons from lattice QCD
have been published in Ref.~\cite{Mathur:2018epb}, in particularly,
$m(B_c^\ast)=6331(4)(6)$~MeV.


In the present work we investigate both radiative and weak nonleptonic decays
of $ B_{(u,d,s,c)}^\ast $ mesons. We use the covariant confined quark model
(CCQM) developed in our previous papers for calculation of the relevant form
factors, branching fractions and decay widths.
Our paper is organized as follows.
In Sec.~\ref{sec:CCQM} we give short sketch to the model
and discuss its basic aspects. 
In Sec.~\ref{sec:elmag}, we present the detailed calculation
of the radiative decays of vector mesons $B^\ast \to B\gamma$ in the
framework of our approach. Then we give the numerical results for
calculated decay widths and compare their values with those obtained in
other approaches. In Sec.~\ref{sec:nonleptonic} we study the weak nonleptonic
decays  $ B^\ast \to D^\ast V $ where $V=\rho,K^\ast,D^\ast,D_s^\ast$.
Since we consider the decays which proceed via tree-level quark diagrams,
the matrix elements of two-body decays are factorized into the leptonic
decays and the weak meson-meson transition. We calculate the relevant
leptonic decay constants and  the form factors of those meson-meson transitions
in the framework of the CCQM. Finally, we compute the two-body
decay widths by using the calculated quantities and compare the obtained
results with other approaches. At the end,  we make a brief summary
of our main results in Sec.~\ref{sec:summary}.

\section{Basic aspects of the covariant confined quark model}
\label{sec:CCQM}

The covariant confined quark model is an effective quantum field approach
to hadronic interactions based on an interaction Lagrangian of hadrons
interacting with their constituent quarks~\cite{Branz:2009cd}.
The coupling strength of the hadrons with the constituent quarks is
determined by the so-called compositeness condition
$Z_H = 0$~\cite{Salam:1962ap,Weinberg:1962hj}  where $Z_H$ is the wave
function renormalization constant of the hadron.
Matrix elements are generated by a set of quark loop diagrams.
The ultraviolet divergences of the quark loops are regularized by
including the hadron-quark vertex functions which, in addition, describe
finite size effects due to the non-pointlike structure of hadrons.
By using Schwinger’s $\alpha$-representation for each local quark propagator
and integrating out the loop momenta, one
can write the resulting matrix element expression as an integral which
includes integrations over a simplex of the
$\alpha$-parameters and an integration over a generalized the Fock-Schwinger
proper time. By introducing an infrared cutoff on the upper limit of
the proper time  one can avoid the appearance of singularities in any matrix
element. The new infrared cutoff parameter $\lambda$  will be taken to
have a common value for all processes.
The CCQM contains only a few model parameters: the light and heavy constituent
quark masses, the size parameters that describe the size
of the distribution of the constituent quarks inside the hadron
and unified parameter $\lambda$.
The model parameters are determined  by a fit to available experimental data. 

The CCQM was successfully applied for description of both light
and heavy hadron exclusive decays. In particularly, the wide range
of the decays of b-hadrons ($B$, $B_s$, $B_c$ and $\Lambda_b$)
have been researched and  described~\cite{Ivanov:2006ni,Ivanov:2000aj,Faessler:2002ut,Gutsche:2015mxa,Ivanov:2005fd,Ivanov:2016qtw,Tran:2018kuv,Ivanov:2002un,Issadykov:2018myx,Dubnicka:2017job,Gutsche:2018utw,Ivanov:2019nqd}.

In this paper we are going to start with an application of the CCQM
to radiative and nonleptonic decays of the vector  excited states
$B^\ast$. An interaction Lagrangian of  $B$-meson which consists
from $\bar{q}b$~quarks ($q=u,d,s,c$) are written as
\be
{\mathcal L}_{\rm int} = g_B\,B(x)J(x) + H.c.
\label{eq:strong-int-Lag}
\en
The local interpolating current has a form $J=\bar q\Gamma b$ where
$\Gamma=i\gamma_5$ for a pseudoscalar meson with spin $S=0$ and
$\Gamma=\gamma^\mu$ for a vector meson with spin $S=1$. The Lorentz index
$\mu$ is contracting with the corresponding index of a vector state.
In the CCQM the nonlocal interpolating currents are used for accounting
the internal structure of a hadron. In our case they  are written as
\be
J(x) =  \int\!\! dx_1 \!\! \int\!\! dx_2 
F_B(x,x_1,x_2) \,\bar q_a(x_1) \, \Gamma \, b_a(x_2)   
\label{eq:cur}
\en
where $a=1,2,3$ is the color index.
The vertex function $F_B(x,x_1,x_2)$ characterizes 
the finite size of the meson.  To satisfy translational invariance the 
vertex function has to obey the identity 
$F_B(x+a,x_1+a,x_2+a) \, = \, F_B(x,x_1,x_2) $
for any given four-vector $a\,$.
We employ a specific form for the vertex function which 
satisfies the translation invariance. One has 
\be
F_B(x,x_1,x_2) \, = \, \delta^{(4)}\left(x - \sum\limits_{i=1}^2 w_i x_i\right)\;  
\Phi_B\biggl( (x_1 - x_2)^2 \biggr) 
\en
where $\Phi_B$ is the correlation function of the two constituent quarks 
with masses $m_{q_1}$ and $m_{q_2}$. The ratios of the quark masses
$w_i$ are defined as 
\be
w_{q_1}=\frac{m_{q_1}}{m_{q_1}+m_{q_2}}, \quad
w_{q_2}=\frac{m_{q_2}}{m_{q_1}+m_{q_2}}, \quad w_1+w_2=1.
\label{eq:w_i}
\en
We choose a simple Gaussian form for the Fourier transform
of vertex function $\widetilde\Phi_B(-k^2)=\exp( k^2/\Lambda_B^2)$.  
The parameter $\Lambda_B$ characterizes the size of the meson. 
Since $k^2$ turns into $-k^2_E$ in  Euclidean space 
the form  $\widetilde\Phi_B(k^2_E)$  has the appropriate fall--off
behavior in the Euclidean region.

The coupling constant $g_B$ in Eq.~(\ref{eq:strong-int-Lag}) is determined
by the so-called {\it compositeness condition}. 
The compositeness condition requires that the renormalization constant $Z_B$ 
of the elementary meson field $B(x)$ is set to zero, i.e.
\be
\label{eq:Z=0}
Z_B=1-\widetilde\Pi^\prime_B(p^2)=0, \qquad (p^2=m^2_B)
\en
where $\Pi^\prime_B(p^2)$ is the derivative of the mass function.

$S$-matrix elements are described by  the quark-loop diagrams
which are the convolution of the vertex functions and quark
propagators. In the evaluation of the quark-loop diagrams we
use the local Dirac propagator
\be
S_q(k) = \frac{1}{ m_q-\not\! k -i\epsilon } = 
\frac{m_q + \not\! k}{m^2_q - k^2  -i\epsilon }
\label{eq:prop}
\en 
with an effective constituent quark mass $m_q$.

The meson functions in the case of the pseudoscalar and vector meson
are written as
\bea
\widetilde\Pi_P(p^2) &=& N_c g_P^2
\int\frac{d^4k}{(2\pi)^4i} \widetilde\Phi^2_P(-k^2)
\Tr\Big(\gamma^5 S_1(k+w_1 p)\gamma^5 S_2(k-w_2 p)\Big),
\label{eq:massP}\\[2ex]
\widetilde\Pi^{\mu\nu}_V(p^2) &=&N_c g_V^2 
\int\frac{d^4k}{(2\pi)^4i} \widetilde\Phi^2_V(-k^2)
\Tr\Big(\gamma^\mu S_1(k+w_1 p)\gamma^\nu S_2(k-w_2 p)\Big)
\nn
&=& g^{\mu\nu} \widetilde\Pi_V(p^2) + p^\mu p^\nu \widetilde\Pi^\parallel_V(p^2).
\label{eq:massV}
\ena
Here  $N_c=3$ is the number of colors.
Since the vector meson is
on its mass-shell  $\epsilon_V\cdot p=0$ we need to keep
the part $ \widetilde\Pi_V(p^2)$. Substituting the derivative of the mass
functions into Eq.~(\ref{eq:Z=0}) one can determine the coupling
constant $g_B$ as a function of other model parameters.
The loop integrations in Eqs.~(\ref{eq:massP}) and ~(\ref{eq:massV})
proceed by using the Fock-Schwinger representation
of quark propagators
\be
S_q (k+w p) = \frac{1}{ m_q-\not\! k- w \not\! p } 
= (m_q + \not\! k + w \not\! p)\int\limits_0^\infty \!\!d\alpha\, 
e^{-\alpha [m_q^2-(k+w p)^2]}.
\label{eq:Fock}
\en
In the obtained integrals over the Fock-Schwinger parameters 
$0\le \alpha_i<\infty$
we introduce an additional integration over the proper time
which converts the set of 
Fock-Schwinger parameters into a simplex. In general case one has
\be
\prod\limits_{i=1}^n\int\limits_0^{\infty} 
\!\! d\alpha_i f(\alpha_1,\ldots,\alpha_n)
=\int\limits_0^{\infty} \!\! dtt^{n-1}
\prod\limits_{i=1}^n \int\!\!d\alpha_i 
\delta\left(1-\sum\limits_{i=1}^n\alpha_i\right)
  f(t\alpha_1,\ldots,t\alpha_n).
\label{eq:simplex}  
\en
Finally, we cut the integration over the proper time
at the upper limit by introducing an infrared cutoff $\lambda$.
One has
\be
\int\limits_0^\infty dt (\ldots) \to \int\limits_0^{1/\lambda^2} dt (\ldots).
\label{eq:conf}
\en
This procedure allows us to remove all possible thresholds present
in the initial quark diagram. Thus the infrared cutoff parameter 
$\lambda$ effectively guarantees the confinement of quarks within hadrons. 
This method is quite general and can be used for diagrams with an arbitrary 
number of loops and propagators. 
In the CCQM the infrared cutoff parameter $\lambda$ is taken to be universal 
for all physical processes.

The model parameters are determined by fitting calculated quantities 
of basic processes to available experimental data or 
lattice simulations (for details, see Ref.~\cite{Branz:2009cd}).
The numerical values of the constituent quark masses
and the cutoff parameter $\lambda$ are given in Table~\ref{tab:fitmas}.
\begin{table}[H]
\caption{Model parameters: quark masses  and cutoff parameter
$\lambda$ (all in GeV).}
\label{tab:fitmas}
\vskip 3mm  
\centering
\def\arraystretch{1.2}
\begin{tabular}{cccc|c}
\hline
$ m_u $  & $ m_s $ & $ m_c  $  &  $  m_b $ & $ \lambda $
\\ \hline
\ \ 0.241\ \   &\ \ 0.428\ \   &\ \ 1.67\ \   &\ \ 5.04\ \ &\ \ 0.181\ \
\\\hline
\end{tabular}
\end{table}


\section{Radiative decays of vector B-mesons}
\label{sec:elmag}


\subsection{{\boldmath$V\to P\gamma$}: theoretical calculation of matrix
  elements and decay widths}

\begin{figure}[H]
\centering
\includegraphics[width=0.7\textwidth]{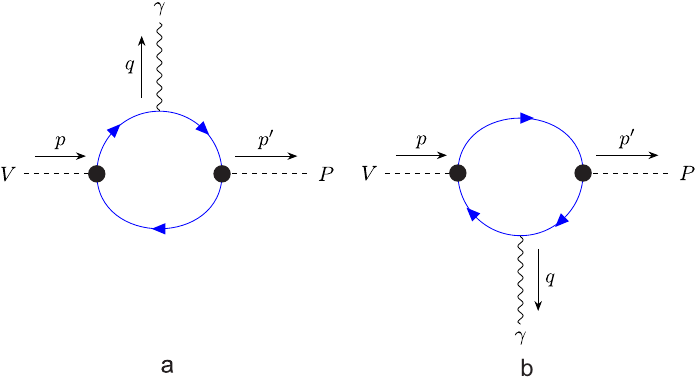}
\caption{Feynman diagrams describing the radiative decays of a vector meson.}
\label{fig:VPg}
 \end{figure}

As the first step we consider the radiative decays of
vector $B^\ast$-mesons. In addition to the strong interaction 
Lagrangian given by Eq.(\ref{eq:strong-int-Lag}), we need the part describing
the electromagnetic interactions.  The free Lagrangian of quarks 
is gauged in the standard manner by using minimal substitution which
gives
\be
\mathcal{L}^{\rm em}_{\rm int}(x) = e\,A_\mu(x)\,J^\mu_{\rm em}(x),\qquad
J^\mu_{\rm em}(x)= e_b\,\bar{b}(x)\gamma^\mu b(x)
+e_q\,\bar{q}(x)\gamma^\mu q(x)
\label{eq:em-lag}
\en
where $e_b$ and $e_q$ are the quark charges in units of the positron charge.
The radiative decays of a vector mesons into a pseudoscalar meson
and photon $V\to P\gamma$ are described by the Feynman diagrams shown
in Fig.~\ref{fig:VPg}.
 One has to note that there is an additional piece in the Lagrangian
 related to the gauging nonlocal interactions of hadrons with
 their constituents~\cite{Branz:2009cd}. This piece gives the additional
 contributions to the electromagnetic processes. However, they are identically
 zero for the process  $V\to P\gamma$ due to its anomalous nature. 

The matrix element of the process $V\to P\gamma$ is written down
\bea
M_{VP\gamma}(p;p',q) = e g_V g_P \epsilon^V_\nu(p)\epsilon^\gamma_\mu(q)
\int\!\! dx\!\!  \int\!\!  dy\!\!  \int\!\!  dz\,
e^{ -ipx + ip^{\prime}y + iqz}
\la\,  T \{ \bar{J}_V^\nu (x)  J^\mu_{\rm em} (z) J_P (y)  \} \ra_0.
\ena
Using the Fourier transforms of the quark currents, we come to the final result
\bea
M_{VP\gamma} (p;p',q) &=& (2\pi)^4 i\, \delta(p-p'-q) M(p,p'),
\nn
M(p, p') &=& (-3i) e g_V g_P \epsilon^V_\nu(p)\epsilon^\gamma_\mu(q)\,
\left( e_b M^{\mu\nu}_b + e_q M^{\mu\nu}_q\right)
\nn
M^{\mu\nu}_b &=&
\int\!\!\frac{dk}{(2\pi)^4 i} \widetilde\Phi_V(-\ell_1^2)
                              \widetilde\Phi_P(-\ell_2^2)
\Tr\left[S_q(k)\gamma^\nu S_b(k-p) \gamma^\mu S_b(k-p')\gamma^5 \right]
\nn
M^{\mu\nu}_q&=& \int\!\!\frac{dk}{(2\pi)^4 i}\widetilde\Phi_V(-\ell_3^2)
                                  \widetilde\Phi_P(-\ell_4^2) 
\Tr\left[S_q(k+p')\gamma^\mu S_q(k+p)\gamma^\nu S_b(k)\gamma^5 \right]
\ena
where $\ell_1=k-w_2\, p$, $\ell_2=k-w_2\, p'$ and
$\ell_3=k+w_1\, p$, $\ell_2=k+w_1\, p'$. The ratios of quark masses
are defined by Eq.~(\ref{eq:w_i}). Now one has
$m_{q_1}=m_b$ and $m_{q_2}=m_q$ with $q=u,d,s$.
By using the technique of calculations outlined in Sec.~\ref{sec:CCQM}
and taking into account the transversality conditions
$\epsilon^\gamma_\mu(q)q^\mu=0$ and $ \epsilon^V_\nu(p)p^\nu=0$
one can arrives at the standard form of matrix element
\be
M(p,p') = e\, g_{V P \gamma}\,\varepsilon^{p q \mu \nu} \epsilon^\gamma_\mu(q)
\epsilon^V_\nu(p),
\en
where $ g_{V P \gamma} =  e_b I_b(m^2_V,m^2_P) + e_q I_q(m^2_V,m^2_P) $
is radiative decay constant. The quantities $I_{b,q}$ are defined by
the two-fold integrals which are calculated numerically.
The electromagnetic decay width is written as
\be
\Gamma(V\to P + \gamma) =
\frac{\alpha}{24} m_V^3\left(1-\frac{m_P^2}{m_V^2}\right)^3 g_{VP\gamma}^2\,.
\en

\subsection{{\boldmath$V\to P\gamma$}: numerical results}

The masses of the $B^\ast$-meson family is not well established yet.
We collect the experimentally measured masses of $B$-meson family
in Table~\ref{tab:B-mass}
\begin{table}[H]
  \caption{Masses of $B$-meson family taken from
    PDG~\cite{ParticleDataGroup:2022pth} (in MeV).}
  \label{tab:B-mass}
\vskip 3mm  
\centering
\def\arraystretch{1.2}
\begin{tabular}{cccc|cc}
\hline
  $B^\pm$ & $B^0$ &  $B^0_s$ &  $B^+_c$ & $B^\ast$ & $B^\ast_s$  \\
\hline
    5279.25(26)  \  &  \ 5279.63(20)  \  &
 \  5366.91(11)  \  &  \ 6274.47(32)  \  &
 \  5324.71(21)  \  &  \ 5415.8(1.5)   \\
\hline
\end{tabular}
\end{table}
%
%

Since the $B_c^\ast$ or $B_c(1^3 S_1)$ state has not been observed yet,
we will use the value $m(B_c^\ast)~=~6331(4)(6)$~MeV obtained from
lattice QCD~\cite{Mathur:2018epb}. The values of size parameters
are taken from our previous papers~\cite{Dubnicka:2017job,Gutsche:2018utw,Ivanov:2019nqd}. Their numerical values are given in Table~\ref{tab:LambdaB}.
\begin{table}[H]
\caption{Size parameters of $B$-meson family (in GeV).}
\label{tab:LambdaB}
\vskip 3mm  
\centering
\def\arraystretch{1.2}
\begin{tabular}{ccc|ccc}
 \hline
$\Lambda_B$ & $\Lambda_{B_s}$ & $\Lambda_{B_c}$ &
$\Lambda_{B^\ast}$ & $\Lambda_{B^\ast_s}$  & $\Lambda_{B^\ast_c}$ \\
\hline 
1.96 \ \  &  \ \ 2.05 \ \  & \ \  2.73 \ \  &  \ \ 1.72 \ \  &  \ \ 1.71 &
\ \ 2.42 \\
\hline
\end{tabular}
\end{table}
The numerical results for the radiative decay constants are shown in
Table~\ref{tab:em-const}. They are compared with those obtained
in Ref.~\cite{Li:2020rcg}.
\begin{table}[H]
\caption{The numerical results for the radiative decay constants (in GeV).}
\label{tab:em-const}
\vskip 3mm  
\centering
\def\arraystretch{1.2}
\begin{tabular}{cccc|c}
\hline
$ g_{B^{\ast\,+} B^+\gamma} $ &  $ g_{B^{\ast\,0} B^0\gamma} $  &
$ g_{B_s^{\ast\,0} B_s^0\gamma} $ & $ g_{B_c^{\ast\,+} B_c^+\gamma} $ & Ref. \\
\hline
\ \ 1.28(13)\ \   &\ \ -0.76(8)\ \  &\ \ -0.57(6)\ \  &\ \ 0.28(3)\ \ &
\ \ CCQM \ \ \\
\hline
\ \ 1.44\ \  &\ \ -0.91 \ \ &\ \ -0.74\ \  & \ \ --- \ \ & \cite{Li:2020rcg}
\\
\hline
\end{tabular}
\end{table}
The numerical values for the radiative decay widths are given
in Tables~\ref{tab:radwidth-1} and  \ref{tab:radwidth-2}.
Our findings are compared with the results obtained in other approaches.

Now, we briefly discuss some error estimates within our model.
The CCQM consists of several free parameters: the constituent quark masses
$m_q$, the hadron size parameters $\Lambda_H$ and the universal infrared
cutoff parameter $\lambda$. These parameters are determined by
minimizing the functional
$\chi^2 = \sum\limits_i\frac{(y_i^{\rm expt}-y_i^{\rm theor})^2}{\sigma^2_i}$
where $\sigma_i$ is the experimental  uncertainty.
If $\sigma$ is too small then we take its value of 10$\%$.
Besides, we have observed that the errors of the fitted parameters 
are of the order of  10$\%$.
Thus, the theoretical error of the CCQM is estimated to be of the order
of 10$\%$ at the level of matrix elements and the order
of 15$\%$ at the level of widths.
\begin{table}[H]
  \caption{Widths of $B^\ast\to B\gamma$ and $B_s^\ast\to B_s\gamma$
    decays (in eV).}
\label{tab:radwidth-1}
\vskip 2mm
\centering
\def\arraystretch{1.2}
\begin{tabular}{c|cccccccc}
\hline
Mode \ & \ CCQM    \ &  \ \cite{Chang:2019xtj} \ & \  
\cite{Choi:2007se} \ & \  \cite{Ebert:2002xz}   \ & \
\cite{Goity:2000dk} \ & \ \cite{Zhu:1996qy}     \ & \
\cite{Ivanov:1994ji} \ & \  \cite{Colangelo:1993zq}
\\
\hline
$B^{\ast +}\to B^+ \gamma$  \ & \ 372(56)  \  &  \ 349(18) \  & \ 400(30) \ & \
190 \ & \ 740(88) \  & \  380(60) \  & \ 401 \  & \  220(90) \\
\hline
$B^{\ast\,0}\to B^0 \gamma $ & 126(19) & 116(6)  & 130(10) & 70 & 228(27) &
130(30) & 131 & 75(27) \\
\hline
$B_s^{\ast\,0}\to B_s^0 \gamma $ & 90(14) & 84(10) &  68(17) & 54 & 136(12) &
220(40) & \\
\hline
\end{tabular}
\end{table}
\begin{table}[H]
\caption{Widths of $B_c^\ast\to B_c\gamma$ decay (in eV).}
\label{tab:radwidth-2}
\vskip 3mm
\centering
\def\arraystretch{1.2}
\begin{tabular}{c|ccccccc}
\hline
Mode \  & \  CCQM  \  &  \ \cite{Liu:2022bdq} \  &  \
\cite{Godfrey:2004ya} \  &  \ \cite{Fulcher:1998ka} \  &  \
\cite{Ebert:2002xz} \  & \ \cite{Gershtein:1994dxw} \  & \
\cite{Eichten:1994gt} \\
\hline
$B_c^{\ast\,+} \to B_c^+ \gamma $ \ & \ 33(5)\ & \ 53(3) \ & \
80 \ & \ 59 \ & \ 33 \ & \ 60 \ & \ 135 \
\\
\hline
\end{tabular}
\end{table}
As can be seen from the Table~\ref{tab:radwidth-1}, there is suppression
of the neutral mode $B^{\ast\,0}\to B^0 \gamma$ compared to the charged mode
$B^{\ast +}\to B^+ \gamma$. The ratio of those widths is written as
\be
\frac{\Gamma(B^{\ast\,0}\to B^0 \gamma)}{\Gamma(B^{\ast +}\to B^+ \gamma)}
\sim \frac{(I_b+I_d)^2}{{(I_b-2I_u)^2}} \approx 0.38.
\label{eq:ratio}
\en
Note that in the heavy quark limit $m_b\to \infty$ the integral $I_b$
is suppressed as $1/m_b$ compare with $I_q$ $(q=u,d)$. Therefore our
result for the ratio in Eq.~(\ref{eq:ratio}) is not far from the
the heavy quark limit $0.25$. This confirms the observation that
the heavy quark limit is quite reliable in the case of $b$-quark.
 

\section{Nonleptonic decays}
\label{sec:nonleptonic}


The effective Hamiltonian describing the nonleptonic decays 
$B^\ast_{u,d,s}\to D^\ast_{u,d,s}\,+\,V$ where $V=D^\ast_{u,d,s}, K^\ast_{u,d},\rho$
is written down
\be
\Heff  = - \frac{G_F}{\sqrt{2}}V^\ast_{cb}V_{q_1q_2}
\Big( C_2
\left( \bar{c}_a O^\mu b_a \right)\left( \bar{q}_{1 b} O_\mu q_{2 b} \right)
+ C_1
\left( \bar{c}_a O^\mu b_b \right)\left( \bar{q}_{1 b} O_\mu q_{2 a}\right)
\Big)
\label{eq:hamiltonian}
\en
where $ G_F $ is the Fermi constant, $V_{cb}$ and $V_{q_1 q_2}$
are the matrix elements of the CKM-matrix, $C_1$ and $ C_2 $ is Wilson
coefficients and $ O^\mu = \gamma^\mu (1-\gamma_5) $ is the weak matrix
with the left chirality. The Wilson coefficients $ C_2 $ (leading order) and
$ C_1 $ (subleading order) are taken at the scale of $ b $-quark mass
from Ref.~\cite{Descotes-Genon:2013vna} (see, Table 3). One has
\be
C_1  = -0.2632 \qquad\text{and}\qquad C_2  = 1.0111.
\en
One has to emphasize that if QCD is neglected, than
$C_1 = 0$ and $ C_2 = 1$  \cite{Buchalla:1995vs,Buras:1998raa}.

\begin{figure}[H]
\centering
\includegraphics[width=0.5\linewidth]{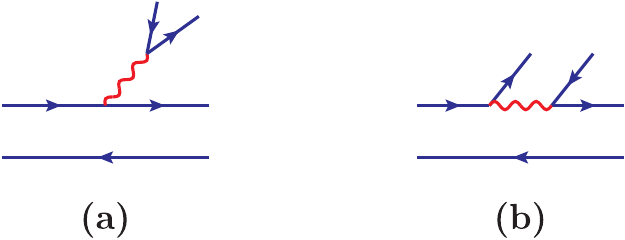}
\caption{Tree-level quark diagrams}
\label{fig:tree}
\end{figure}

\begin{figure}[H]
  \centering
    \includegraphics[width=0.3\textwidth]{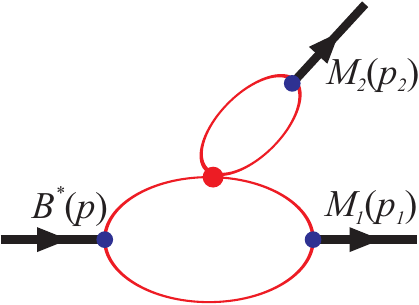}
\caption{Diagram describing the nonleptonic decay $ B^\ast \to M_1 M_2 $.}
\label{fig:BV1V2_CCQM}
\end{figure}

We consider the nonleptonic decays $B^\ast\to D^\ast V$ which proceed via
tree-level quark diagrams shown in Fig.~\ref{fig:tree}.
One has to note that the Wilson coefficients will appear in
combinations either $a_1=C_2+\xi C_1$ for the charged emitted meson or
$a_2=C_1+\xi C_2$  for the neutral emitted meson.
The color suppression factor  $ \xi=1/N_c $ will be used to be equal
to zero in calculation that corresponds to the
large-$ N_c $ limit $ \xi=0$. One has to remind that an
approximation is widely used in the phenomenological studies
of two-body nonleptonic decays because in the case of $N_c=3$
the combination $a_2\approx 0.074$, i.e. significantly suppressed.
The Feynman diagram describing the nonleptonic decays $B^\ast\to M_1M_2$
within the CCQM are shown in Fig.~\ref{fig:BV1V2_CCQM}.
The matrix element of the nonleptonic $ B^\ast\to M_1(p_1)M_2(p_2) $ decays are
given by
\bea
\label{eq:matrix-elem}
M(B^\ast(p)  \to M_1(p_1) M_2(p_2)) &=&
\frac{G_F}{\sqrt{2}}V_{CKM}a_W\,
m_2f_{M_2}\epsilon_{2\,\mu}\,
\nn
&\times&\Bla M_1\left(p_1,\epsilon_{1\,\beta}\right) |\bar q_1 O^\mu q_2 |
B^\ast(p,\epsilon^\ast_{\alpha})\Bra
\ena
where $f_{M_2}$ is the leptonic decay constant of the vector  meson $M_2$.
Here $V_{CKM}$ and $a_W$ are the relevant CKM-matrix elements and
the Wilson coefficients, respectively.
  Note that the Eq.~(\ref{eq:matrix-elem}) is equivalent to the factorization
  hypothesis. 

The leptonic decay constant of the vector meson  $M_2$ which consists from
$q_1$ and $q_2$ quarks is defined by
\be
N_c\, g_{M_2}\! \int\!\! \frac{d^4k}{ (2\pi)^4 i}\, \widetilde\Phi_{M_2}(-k^2)\,
\Tr \biggl[O^{\,\mu} S_1(k+w_1 p_2)\not\!\epsilon_2  S_2(k-w_2 p_2) \biggr] 
= m_2 f_{M_2} \epsilon_2^\mu.
\label{eq:lept}
\en
The meson is taken on its mass-shell, i.e.
$p_2^2=m^2_2$ and $\epsilon_2\cdot p_2=0$.

The matrix elements of the weak $B^\ast-M_1$ transition
is written as
\begin{align}
 & \la M_1(p_1) | \bar{q}_1   O^\mu q_2  | B^\ast(p) \ra  =
  \nn
  & = N_c\, g_{B^\ast}\,g_{M_1}\epsilon_\alpha(p)\epsilon^\ast_{1\,\beta}(p_1)
  \!\!  \int\!\! \frac{d^4k}{ (2\pi)^4 i}\, 
\widetilde\Phi_{B^\ast}\Big(-(k+w_{13} p_1)^2\Big)\,
\widetilde\Phi_{M_1}\Big(-(k+w_{23} p_2)^2\Big)
\nn
& \times \Tr\biggl[
O^{\,\mu}\, S_1(k+p_1)\,\gamma^\alpha\, S_3(k)\,\gamma^\beta \,S_2(k+p_2)\biggr].   
\label{eq:V1V2}
\end{align}
All particles are on their mass shell:
$p^2=m^2_{B^\ast}$, $\epsilon_\alpha\, p^\alpha =0$, and
$p_1^2=m^2_1$, $\epsilon_{1\,\beta}\, p_1^\beta =0$.
Altogether there are three flavors of quarks involved in this
process. We therefore introduce a notation with two subscripts
$w_{ij}=m_{q_j}/(m_{q_i}+m_{q_j})$ $(i,j=1,2,3)$ such that $w_{ij}+w_{ji}=1$. 
The loop integrations in the Eqs.~(\ref{eq:lept}) and (\ref{eq:V1V2})
are performed by using technique given in Sec.~\ref{sec:CCQM}.
We finalize by two- and three-fold integrals in equations 
Eqs.~(\ref{eq:lept}) and (\ref{eq:V1V2}), respectively. We calculate them
by using FORTRAN codes with NAG library.

The matrix elements of the weak $B^\ast-M_1$ transition
can be expressed in terms of six vector form factors $V_i$
and four axial form factors $A_i$. One has
\begin{align}
\la M_1(p_1) | \bar{q}_1 & O^\mu q_2  | B^\ast(p) \ra  =
\epsilon_\alpha\epsilon^\ast_{1\,\beta}\times
\Big[
g^{\alpha\beta} \left(- P^\mu  V_1(q^2) + q^\mu V_2 (q^2) \right)
\nn
& + \frac{q^\alpha q^\beta}{M^2 - m^2_1}
   \left( P^\mu V_3(q^2) - q^\mu V_4(q^2)\right) 
   - q^\alpha g^{\mu\beta} V_5 (q^2) + q^\beta g^{\mu\alpha} V_6 (q^2)
     \Big]
     \nn[1.5ex]
 &    
+\epsilon_\alpha\epsilon^\ast_{1\,\beta}\times
\Big[    
- i \varepsilon^{\mu \nu \alpha \beta}
\left( P^\nu A_1 (q^2) - q^\nu A_2 (q^2) \right)
\nn
& + \frac{ i \varepsilon^{\mu \nu q P}} {M^2 - m^2_1}
\left( g^{\nu\alpha}q^\beta A_3 (q^2) - g^{\nu\beta} q^\alpha A_4 (q^2) \right)
\Big]
\label{eq:formfactor}
\end{align}
where $P=p+p_1$, $ q = p - p_1 = p_2 $.
The full matrix element of the nonleptonic decay
$B^\ast(p)\to M_1(p_1) M_2(p_2)$ is obtained by contraction of the above
two expressions from Eqs.~(\ref{eq:lept}) and (\ref{eq:formfactor}).
Keeping in mind that
$ \epsilon^\ast_2 \cdot q = (\epsilon^\ast_2 \cdot p_2) = 0 $ 
one gets that the two vector form factors  $ V_2 $ and $ V_4 $
do not contribute to the full matrix element.

The values of the size parameters of vector mesons are taken from
Ref.~\cite{Dubnicka:2017job,Gutsche:2018utw,Ivanov:2019nqd} and displayed in
Table~\ref{tab:vectorsize}.
\begin{table}[H]
\caption{Size parameters of vector mesons (in GeV)}
\label{tab:vectorsize}
\vskip 2mm
\centering
\def\arraystretch{1.2}
\begin{tabular}{cccccc}
\hline
\ \ $\Lambda_{\rho}$ \ \  &  \ \  $\Lambda_{K^\ast}$  \ \ &  \ \
$\Lambda_{D^\ast}$ \ \  & \ \  $\Lambda_{D^\ast_s}$ \ \  &  \ \
$\Lambda_{B^\ast}$  \ \ &  \ \ $\Lambda_{B^\ast_s}$ \\
\hline
0.61  & 0.81 & 1.53 & 1.56 & 1.72 & 1.71
\\
\hline
\end{tabular}
\end{table}

The calculated leptonic decay constants  $f_{M_2}$ are shown in
Table~\ref{tab:lepconst}.
\begin{table}[H]
\caption{Calculated leptonic decay constants  $f_{M_2}$ (in MeV)}
\label{tab:lepconst}
\vskip 2mm
\centering
\def\arraystretch{1.1}
\begin{tabular}{ccl}
\hline
 \ \  & \ \  CCQM  \ \  & \ \  Expt/Lat  \\
\hline
$f_\rho$       & 218(22) & $221(1) \quad $ \ \cite{ParticleDataGroup:2022pth}
\\
$f_{K^\ast}$    & 227(23) & $217(7) \quad $  \ \cite{ParticleDataGroup:2022pth}
\\
$f_{D^\ast}$    & 246(25) & $223.5(8.4)$~\cite{Lubicz:2017asp}
\\
$f_{D^\ast_s}$  & 273(27) & $268.8(6.6)$~\cite{Lubicz:2017asp}
\\
$f_{B^\ast}$   & 185(19) & $186.4(7.1)$ \cite{Lubicz:2017asp,ETM:2016nbo}
\\
$f_{B^\ast_s}$   & 260(26) & $223.1(5.6)$~\cite{Lubicz:2017asp,ETM:2016nbo}
\\
\hline
\end{tabular}
\end{table}

\begin{figure}[H]
\centering
\begin{minipage}[b]{0.45\linewidth}
\includegraphics[width=\linewidth]{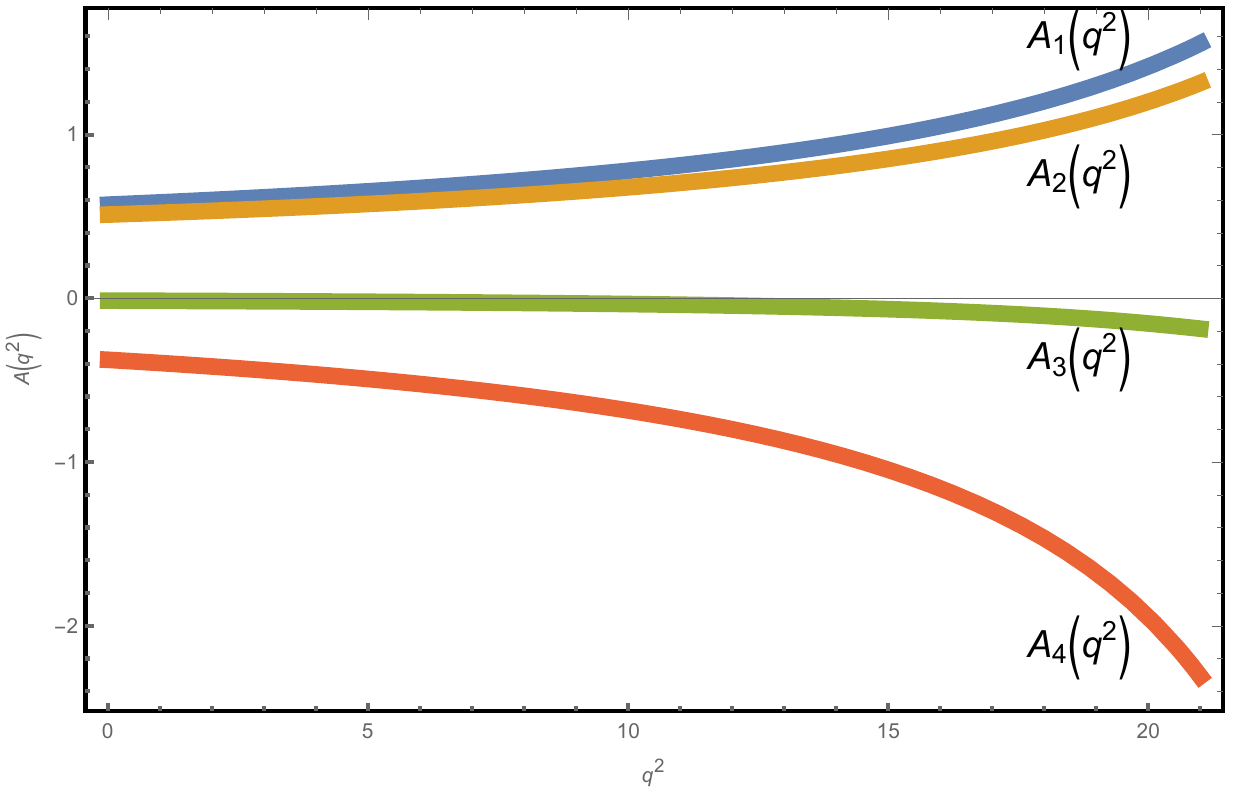}
\end{minipage}
\begin{minipage}[b]{0.45\linewidth}
\includegraphics[width=\linewidth]{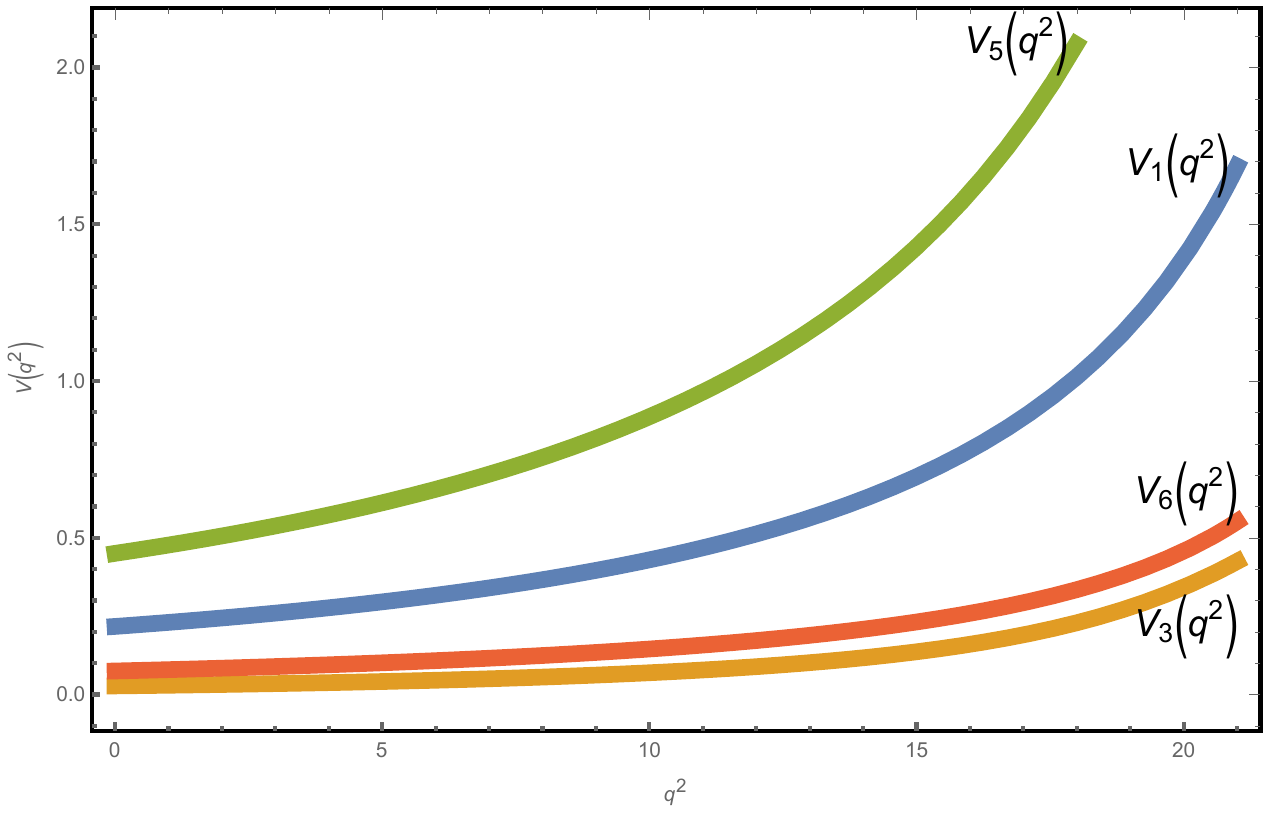}
\end{minipage}
\caption{Form factors of $ B^\ast-\rho $~transition}
\label{fig3:brho}
\end{figure}

\begin{figure}[H]
\centering
\begin{minipage}[b]{0.45\linewidth}
\includegraphics[width=\linewidth]{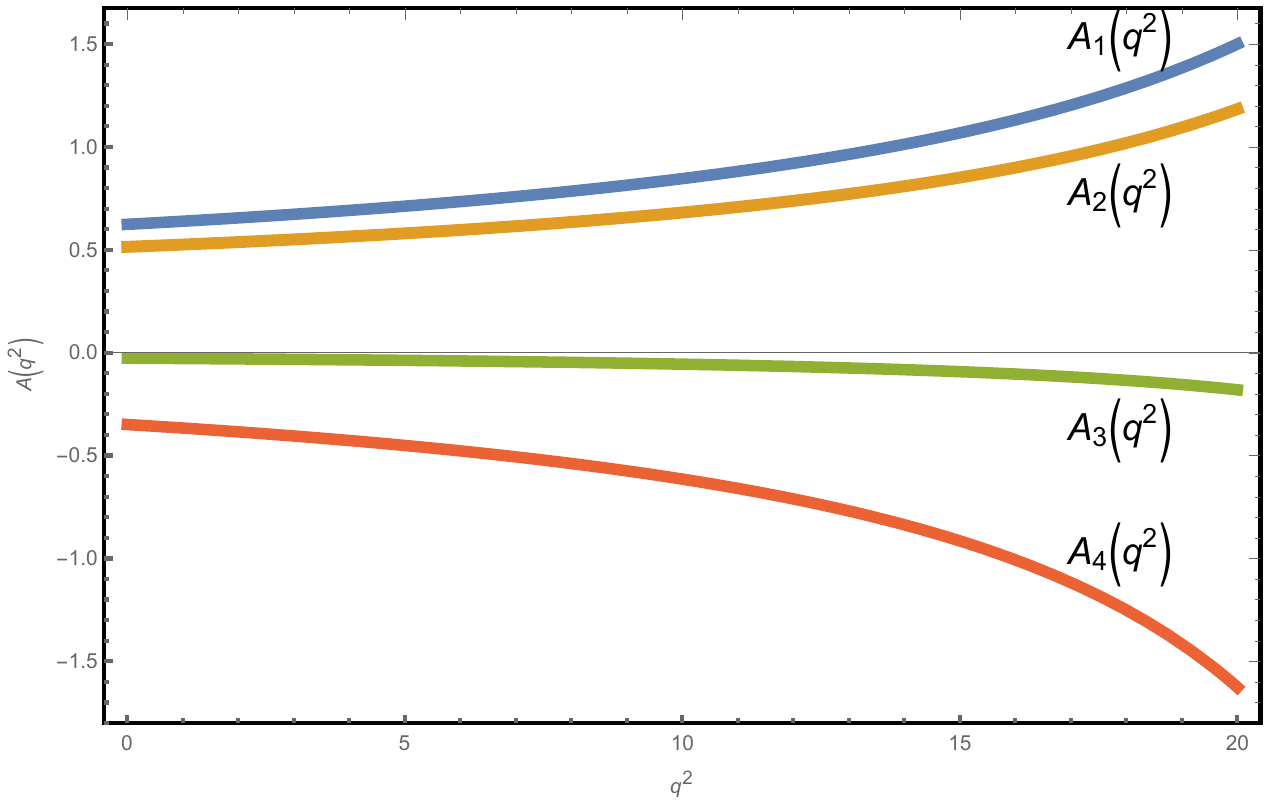}
\end{minipage}
\begin{minipage}[b]{0.45\linewidth}
\includegraphics[width=\linewidth]{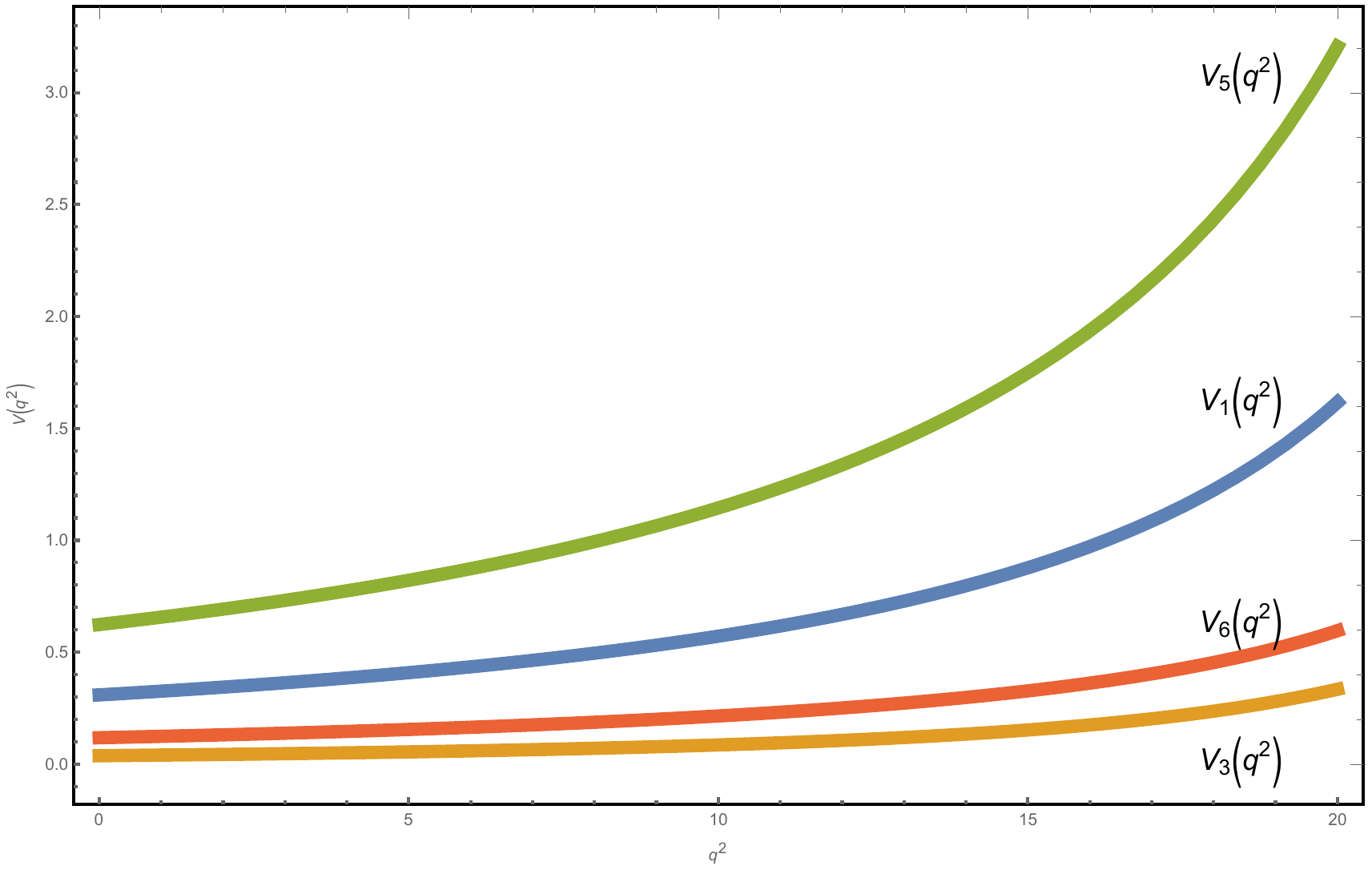}
\end{minipage}
\caption{Form factors of $ B^\ast-K^\ast $~transition}
\label{fig4:bd}
\end{figure}

\begin{figure}[H]
\centering
\begin{minipage}[b]{0.45\linewidth}
\includegraphics[width=\linewidth]{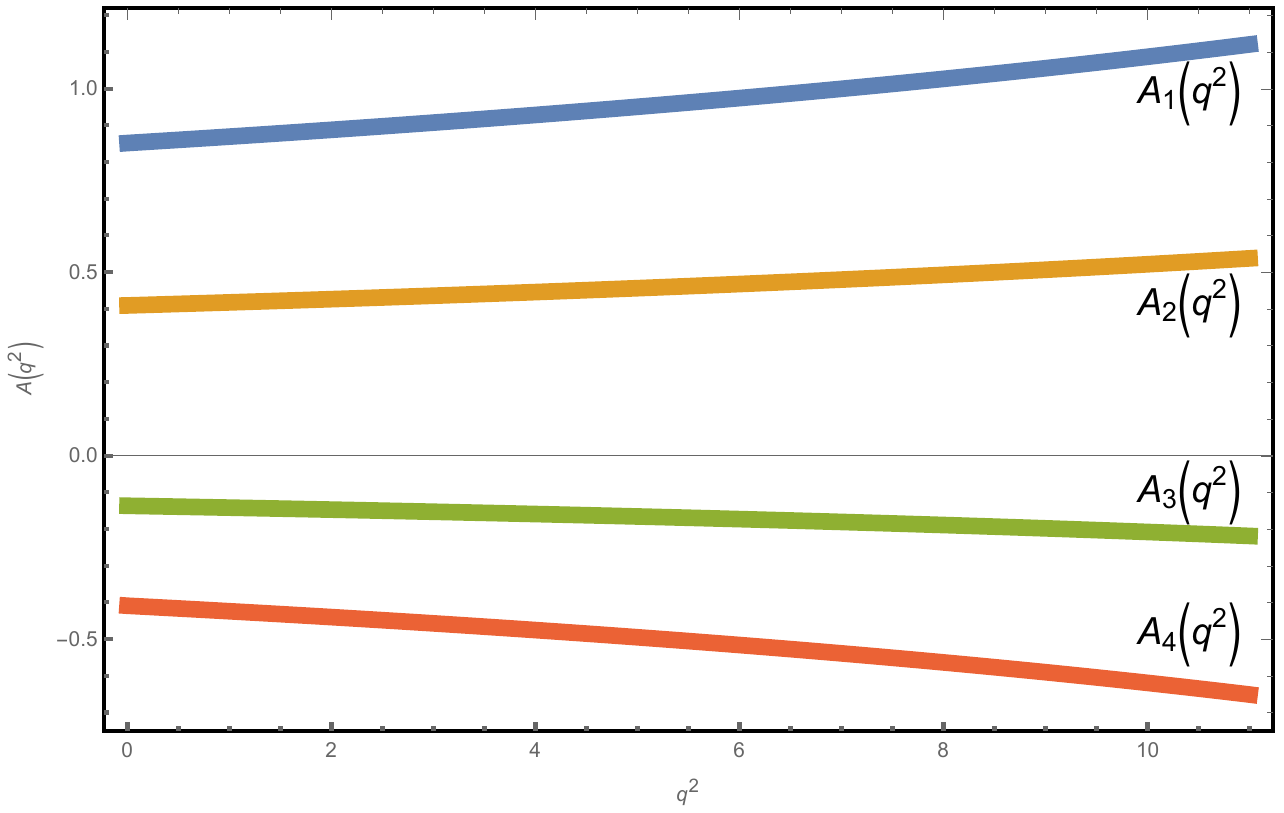}
\end{minipage}
\begin{minipage}[b]{0.45\linewidth}
\includegraphics[width=\linewidth]{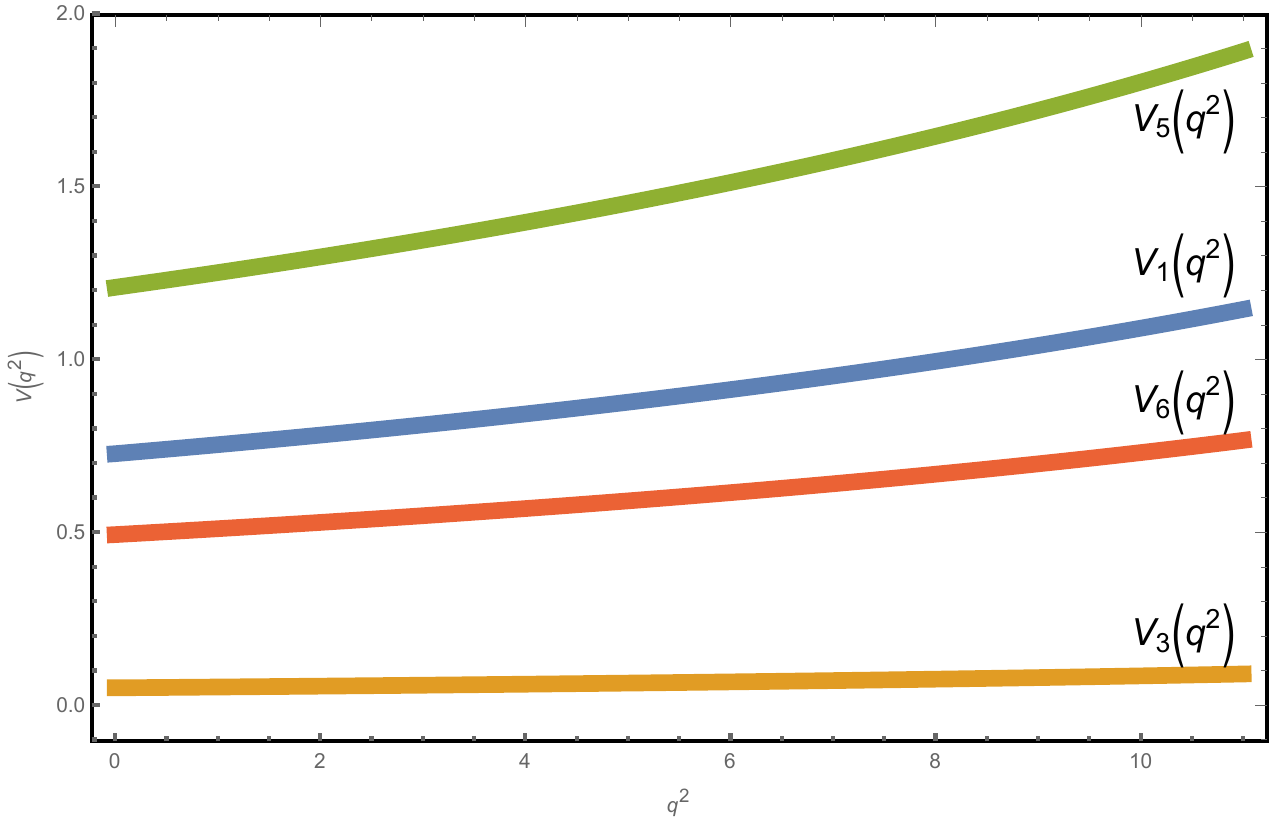}
\end{minipage}
\caption{Form factors of $ B^\ast-D^\ast $~transition}
\label{fig5:bd}
\end{figure}

\begin{figure}[H]
\centering
\begin{minipage}[b]{0.45\linewidth}
\includegraphics[width=\linewidth]{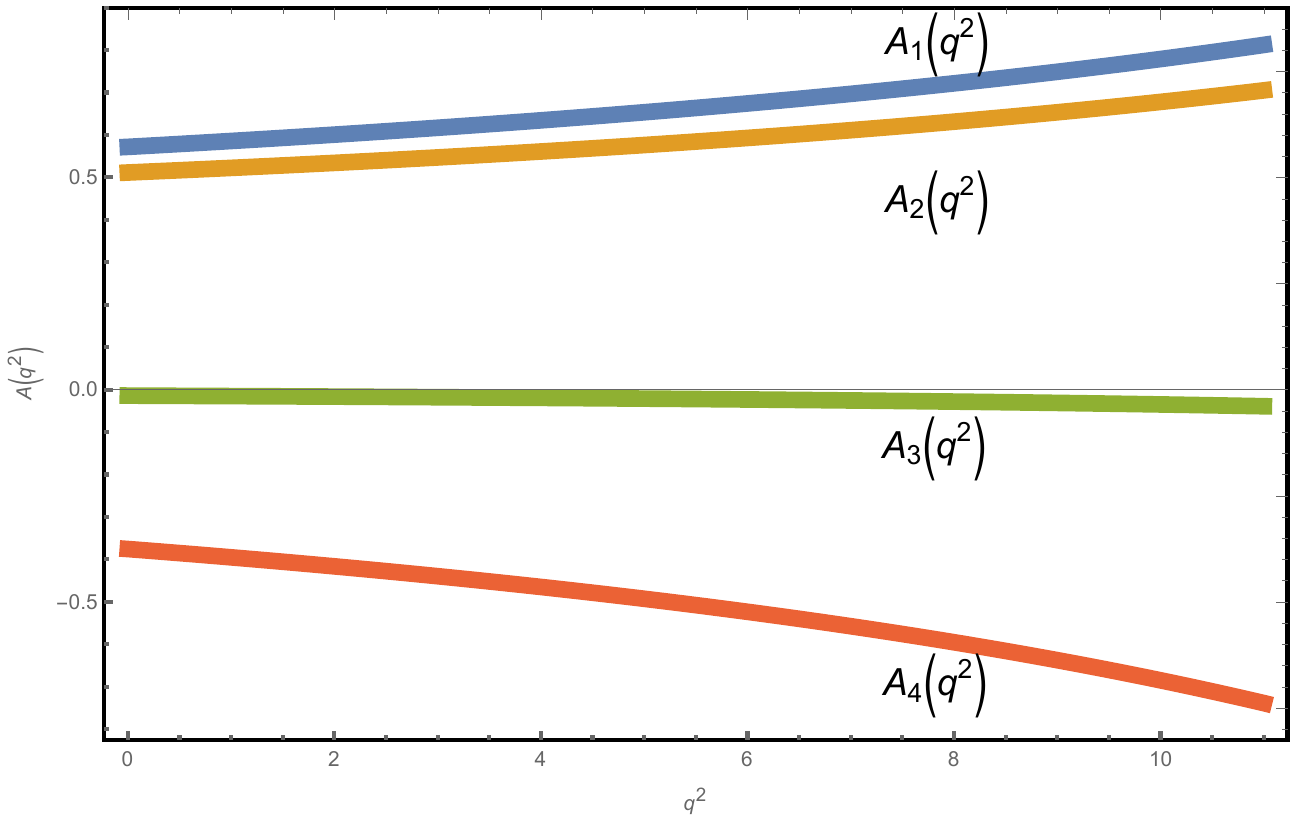}
\end{minipage}
\begin{minipage}[b]{0.45\linewidth}
\includegraphics[width=\linewidth]{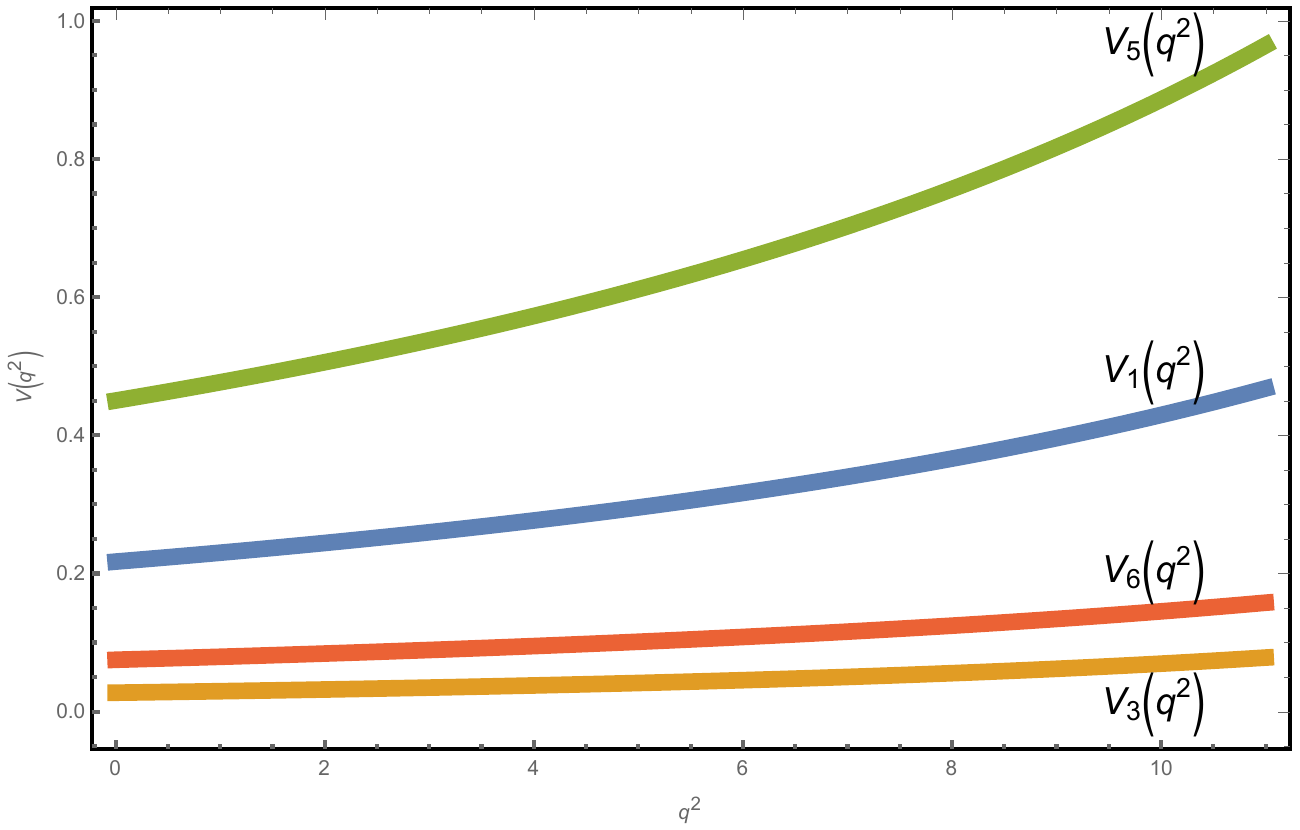}
\end{minipage}
\caption{Form factors of $ B_s^\ast-D_s^\ast $~transition}
\label{fig6:bd}
\end{figure}

We calculate the relevant transition form factors in the full kinematical
region of the transfered momentum squared $q^2$. The behavior
of the form factors on the $q^2$ are shown in
Figs.~\ref{fig3:brho}-\ref{fig6:bd}.

The numerical results for the form factors are well approximated 
by a dipole parametrization
\be
F(q^2) = \frac{F(0)}{1-a s + b s^2}, \qquad s=\frac{q^2}{M^2}
\label{eq:dipole}
\en
where $M$ is the mass of ingoing meson. The dipole approximation is quite
accurate. The error relative to the exact results is less than 1$\%$ over
the entire $q^2$ range.


\begin{table}[H]
 \caption{Parameters of dipole approximation }
\label{tab:dipole}
\vskip 2mm
\centering
\def\arraystretch{1.1} 
\begin{tabular}{c|cccc|cccc}
\hline  
\ \ $B^\ast-\rho$ \ \ & \ \
$A_1$ \ \ & \ \  $A_2$ \ \ & \ \ $A_3$ \ \  & \ \ $A_4$  \ \ & \ \
$V_1$ \ \ & \ \  $V_3$ \ \ & \ \ $V_5$ \ \  & \ \ $V_6$ \ \  
\\
\hline
\ \ F(0) \ \  & \ \
0.57 \ \  & \ \ 0.51  \ \  & \ \ $-0.014$ \ \ & \ \ $-0.38$ \ \ & \ \   
0.22 \ \  & \ \ 0.027 \ \  & \ \   0.45   \ \ & \ \  0.075 \ \ 
\\ 
a    & 0.66 & 0.58 & 2.08 & 1.42               
     & 1.61 & 2.13 & 1.59 & 1.55
\\
b     & $-0.26$ & $-0.33$ & 1.13 & 0.39
      & 0.58 & 1.17 & 0.57 & 0.52
\\ 
\hline
\end{tabular}\\[2ex]
\begin{tabular}{c|cccc|cccc}
\hline  
\ \ $B^\ast-K^\ast$\ \  &\ \
$A_1$\ \ &\ \ $A_2$\ \  &\ \ $A_3$\ \ &\ \  $A_4$ \ \ &\ \ 
$V_1$\ \ &\ \ $V_3$\ \  &\ \ $V_5$\ \ &\ \ $V_6$  \ \  
\\
\hline
\ \ F(0)\ \ &\ \
$0.62$\ \   &\ \ $0.51$\ \ & \ \ $-0.027$\ \ &\ \ $-0.35$\ \ & \ \
$0.31$ \ \  & \ \  $0.038$\ \  & \ \ $0.62$ \ \ & $0.12$ \ \ 
\\ 
a & $0.66$ & $0.59$ & $1.75$ & $1.34$
  & $1.46$ & $1.92$ & $1.44$ & $1.42$
\\
b & $-0.24$ & $-0.31$  & $0.77$ &  $0.31$
          & $0.44$    & $0.94$    & $0.42$ &  $0.40$
\\ 
\hline
\end{tabular}\\[2ex]
\begin{tabular}{c|cccc|cccc}
\hline  
$B^\ast-D^\ast$\ \ &\ \ $A_1$\ \ &\ \  $A_2$\ \ &\ \ $A_3$\ \  &\ \ $A_4$  \ \ 
                 &\ \ $V_1$\ \ &\ \ $V_3$\ \ &\ \ $V_5$\ \ &\ \ $V_6\qquad$  
\\
\hline
F(0)\ \  & \ \ 0.85\ \  & \ \ 0.41 \ \  & \ \ $-0.14$ \ \ & $-0.41$\ \  
         & \ \ 0.73\ \  & \ \ 0.050 \ \ & \ \   1.21\ \   &\ \  0.49\ \ 
\\ 
a   & 0.57 & 0.56 & 1.02 & 1.01
    & 0.98 & 1.29 & 0.97 & 0.95
\\
b     & $-0.14$ & $-0.15$ & 0.14 & 0.13
      & $0.11$ & 0.36 & 0.10 & 0.089
\\ 
\hline
\end{tabular}\\[2ex]
\begin{tabular}{c|cccc|cccc}
\hline  
$B_s^\ast-D_s^\ast$\ \ &\ \ $A_1$\ \ &\ \  $A_2$\ \ &\ \ $A_3$\ \  &\ \ $A_4$\ \ 
                &\ \ $V_1$\ \ &\ \ $V_3$\ \ &\ \ $V_5$ \ \ &\ \ $V_6\qquad$  
\\
\hline
F(0)\ \ & \ \ 0.78\ \  & \ \ 0.38 \ \  & \ \ $-0.11$ \ \ & $-0.34$\ \  
        & \ \ 0.68\ \  & \ \ 0.051 \ \ & \ \   1.16\ \   &\ \  0.49\ \ 
\\ 
a     & 0.71 & 0.69 & 1.16 & 1.11
      & 1.10 & 1.36 & 1.08 & 1.07
\\
b     & $-0.13$ & $-0.16$ & 0.21 & 0.15
      & $0.15$ & 0.38 & 0.15 & 0.13
\\ 
\hline
\end{tabular}
\end{table}


\begin{table}[H]
  \caption{Form factors at origin $q^2=0$ compared
    with CLFQM~\cite{Chang:2019xtj}  }
	\label{tab:comparison}
\vskip 2mm
	\centering
	\begin{tabular}{|c|c|c|}
		\hline
\ \	 $ B^\ast-\rho $ \ \ & \ \ CCQM \ \  & \ \ CLFQM \ \  \\
		\hline
		$A_1$   & 0.57 & 0.27 \\
		\hline
		$A_2$   & 0.51 & 0.25 \\
		\hline
		$A_3$   & -0.014 & 0.07 \\
		\hline
		$A_4$   & -0.38 & 0.06  \\
		\hline
		$V_1$   & 0.22 & 0.28  \\
		\hline
		$V_3$   & 0.027 & 0.11  \\
		\hline
		$V_5$   & 0.45 & 0.60 \\
		\hline
		$V_6$   & 0.075 & 0.14  \\
		\hline
	\end{tabular}
\hspace*{1cm}	\begin{tabular}{|c|c|c|}
		\hline
 \ \ 	$B^\ast - K^\ast$ \ \ 	& \ \  CCQM \ \   &  \ \  CLFQM \ \     \\
		\hline
		$A_1$   & 0.62 & 0.33 \\
		\hline
		$A_2$   & 0.51 & 0.27 \\
		\hline
		$A_3$   & -0.027 & 0.07 \\
		\hline
		$A_4$   & -0.35 & 0.07 \\
		\hline
		$V_1$   & 0.31 & 0.33 \\
		\hline
		$V_3$   & 0.038 & 0.11 \\
		\hline
		$V_5$   & 0.62 & 0.68 \\
		\hline
		$V_6$   & 0.12 & 0.16 \\
		\hline
	\end{tabular}\\[2ex]
	\begin{tabular}{|c|c|c|}
		\hline
 \ \    $B^\ast - D^\ast$ \ \  & \ \  CCQM  \ \   & \ \  CLFQM  \ \ \\
		\hline
		$A_1$   & 0.85 & 0.66 \\
		\hline
		$A_2$   & 0.41 & 0.35 \\
		\hline
		$A_3$   & -0.14 & 0.07 \\
		\hline
		$A_4$   & -0.41 & 0.08 \\
		\hline
		$V_1$   & 0.73 & 0.67 \\
		\hline
		$V_3$   & 0.050 & 0.13 \\
		\hline
		$V_5$   & 1.21 & 1.17 \\
		\hline
		$V_6$   & 0.49 & 0.48 \\
		\hline
	\end{tabular}
\hspace*{1cm}	\begin{tabular}{|c|c|c|}
		\hline
 \ \   $B^\ast_s \to D^\ast_s$ \ \ 	&  \ \ CCQM \ \   &  \ \ CLFQM \ \   \\
		\hline
		$A_1$   & 0.78 & 0.65 \\
		\hline
		$A_2$   & 0.38 & 0.38 \\
		\hline
		$A_3$   & -0.11 & 0.10 \\
		\hline
		$A_4$   & -0.34 & 0.09 \\
		\hline
		$V_1$   & 0.68 & 0.66 \\
		\hline
		$V_3$   & 0.051 & 0.15 \\
		\hline
		$V_5$   & 1.16 & 1.19 \\
		\hline
		$V_6$   & 0.49 & 0.53 \\
		\hline
	\end{tabular}
\end{table}

For convenience of calculation,
we further introduce the helicity amplitudes
$H^{M_1M_2}_{\lambda_1\lambda_2}$ for the outgoing mesons $M_1$
and $M_2$. We imply that the meson $M_2$ is the emitted meson
whereas $M_1$ is the outgoing meson in the transition~$B^\ast-M_1$.
One has
\begin{align}
  H^{M_1M_2}_{00} =f_{M_2}m_2 & \Big[
  + \frac{|\mathbf{p_2}| \left( M^2 + m_1^2 - m_2^2 \right)}{m_1 m_2} V_1
  + \frac{2 M^2 |\mathbf{p_2}|^3}{\left( M^2 - m_1^2\right)m_1 m_2} V_3
\nn      
 &\,\,- \frac{|\mathbf{p_2}|\left( M^2 - m_1^2 - m_2^2\right)}{2 m_1 m_2}V_5
  + \frac{|\mathbf{p_2}|\left( M^2 - m_1^2 + m_2^2\right)}{2 m_1 m_2}V_6\Big]\,,
\nn[1.3ex]
H^{M_1M_2}_{++} = f_{M_2}m_2 & \Big[+\frac{3 M^2 + m_1^2 - m_2^2}{2 M} A_1
- \frac{M^2-m_1^2+m_2^2}{2M} A_2 + \frac{2|\mathbf{p_2}|^2 M}{M^2-m_1^2} A_4
- |\mathbf{p_2}| V_5\Big]\,,
\nn[1.3ex]
H^{M_1M_2}_{--} = f_{M_2}m_2 & \Big[-\frac{3 M^2 + m_1^2 -m_2^2}{2M}A_1
          + \frac{M^2-m_1^2+m_2^2}{2M}A_2
          - \frac{2|\mathbf{p_2}|^2 M}{M^2 - m_1^2} A_4
          - |\mathbf{p_2}| V_5\Big]\, ,
\nn[1.3ex]
H^{M_1M_2}_{+0} = f_{M_2}m_2 & \Big[-\frac{M^2-m_1^2}{m_2}A_1 + m_2 A_2
+ \frac{2 M |\mathbf{p_2}|}{m_2} V_1\Big]\,,
\nn[1.3ex]
H^{M_1M_2}_{-0}  = f_{M_2}m_2 & \Big[+ \frac{M^2-m_1^2}{m_2}A_1 - m_2 A_2
  + \frac{2 M |\mathbf{p_2}|}{m_2} V_1\Big]\, ,
\nn[1.3ex]
H^{M_1M_2}_{0-} = f_{M_2}m_2 & \Big[-\frac{M^2+3m_1^2-m_2^2}{2m_1}A_1
  + \frac{M^2 -m_1^2 -m_2^2}{2m_1} A_2
  - \frac{2 |\mathbf{p_2}|^2 M^2}{\left(M^2-m_1^2\right)m_1} A_3
\nn  
  &\,\,-  \frac{M |\mathbf{p_2}|}{m_1} V_6\Big]\,,
\nn[1.3ex]
H^{M_1M_2}_{0+} =  f_{M_2}m_2 & \Big[+\frac{M^2+3m_1^2-m_2^2}{2m_1}A_1
  - \frac{M^2 -m_1^2 -m_2^2}{2m_1} A_2
  + \frac{2 |\mathbf{p_2}|^2 M^2}{\left(M^2-m_1^2\right)m_1} A_3
\nn
  &\,\,-  \frac{M |\mathbf{p_2}|}{m_1} V_6\Big]\,.
\nonumber
\end{align}
Here $ |\mathbf{p_2}| = \lambda^{1/2} (M^2,m_1^2,m_2^2)/2M$ is
the momentum of the daughter meson in the $B^\ast$-meson rest frame.

Finally, the amplitudes can be written in terms of helicity amplitudes.
The only two amplitudes contain the contributions from the diagrams
with the charged and neutral emitted mesons. They are written as
\bea
M(B^{\ast -} \to D^{\ast 0} + K^{\ast -}) &=&
\frac{G_F}{\sqrt{2}}
\Big( a_1 V_{c b}V^\ast_{u s} H_{\lambda_{D^\ast}\lambda_{K^\ast}}^{D^\ast K^\ast} 
     +a_2  V_{c b}V^\ast_{u s} H_{\lambda_{K^\ast}\lambda_{ D^\ast}}^{K^\ast D^\ast}  \Big), 
     \nn
M(B^{*-} \to D^{*0} + \rho^{-}) &=&
\frac{G_F}{\sqrt{2}}
\Big(  a_1 V_{c b}V^\ast_{u d} H_{\lambda_{D^\ast}\lambda_{\rho}}^{D^\ast \rho} 
     + a_2  V_{c b}V^\ast_{u d} H_{\lambda_{\rho}\lambda_{D^\ast}}^{\rho D^\ast}  \Big).
     \label{eq:ampl-Krho}
\ena     
Other ten amplitudes contain the contributions from the diagrams
with the charged  emitted mesons only. One has
\begin{table}[H]
  \caption{Amplitudes of  $B^\ast\to D^\ast V_{\rm charge}$ decays.}
\label{tab:ampl-charge}
\vskip 2mm
\def\arraystretch{1.1}
\centering
\begin{tabular}{|c|c||c|c|}
  \hline
\multicolumn{4}{|c|}{$M(B^\ast\to D^\ast V_{\rm charge}) =
  {\bf \frac{G_F}{\sqrt{2}} a_1V_{cb}}\,
  \widetilde{M}(B^\ast\to D^\ast V_{\rm charge})$}
\\[1.2ex]
\hline
Mode & $\widetilde{M}$ &  Mode & $\widetilde{M}$ \\
\hline
 \ \  $B^{*-} \to D^{*0}\,D^{*-}$ \ \   & \ \  
 $ V^\ast_{c d}\,  H_{\lambda_{D^{\ast 0}} \lambda_{D^{\ast -}}}^{D^{\ast 0} D^{\ast -}} $\ \  &\ \
$  B^{*-} \to D^{*0}\,D^{*-}_s$  \ \  &  \ \  
$ V^\ast_{c s}\,  H_{\lambda_{D^{\ast}}\lambda_{D_s^{\ast}}}^{D^\ast D^\ast_s} $  \ \  
 \\
 $ \bar{B}^{*0}\to D^{*+}\,K^{*-} $
 & $V^\ast_{u s}\,H_{\lambda_{D^{\ast}}\lambda_{K^{\ast}}}^{D^\ast K^\ast} $ &
 $ \bar{B}^{*0}\to D^{*+}\,\rho^{-} $ &
 $ V^\ast_{u d}\,H_{\lambda_{D^{\ast}}\lambda_{\rho}}^{D^\ast \rho} $
 \\
 $\bar{B}^{*0}\to D^{*+}\,D^{*-}$ &
 $V^\ast_{c d}\, H_{\lambda_{D^{\ast +}} \lambda_{D^{\ast -}}}^{D^{\ast +} D^{\ast -}} $ &
 $\bar{B}^{*0}\to D^{*+}\,D_s^{*-} $ &
 $ V^\ast_{c s}  H_{\lambda_{D^{\ast}}\lambda_{D^{\ast}_s}}^{D^\ast D^\ast_s}$
 \\
 $\bar{B}_s^{*0}\to D^{*+}_s\,K^{*-} $ &
 $V^\ast_{u s}\,  H_{\lambda_{D_s^{\ast}}\lambda_{K^{\ast }}}^{D^\ast_s K^\ast} $  &
 $\bar{B}_s^{*0}\to D^{*+}_s\,\rho^{-} $ &
 $V^\ast_{u d}\,  H_{\lambda_{D_s^{\ast}} \lambda_{\rho}}^{D^\ast_s \rho} $
 \\
 $\bar{B}_s^{*0}\to D^{*+}_s\,D^{\ast -} $ &
 $V^\ast_{c d} \, H_{\lambda_{D_s^{\ast}}\lambda_{D^{\ast}}}^{D^\ast_s D^\ast} $ &
 $\bar{B}_s^{*0}\to  D^{*+}_s \,D^{\ast -}_s $ &
 $V^\ast_{c s}\,  H_{\lambda_{D_s^{\ast +}} \lambda_{D_s^{\ast -}}}^{D^{\ast +}_s D^{\ast -}_s} $
\\
 \hline 
\end{tabular}
\end{table}
By assuming that the $B^\ast$-meson is unpolarized, the decay widths
are calculated by the formulas
\be
\Gamma\left(B^{\ast\,\lambda} \to M_1^{\lambda_1} M_2^{\lambda_2}\right) = 
\frac{|\mathbf{p_2}|}{24\pi\,M^2}
\sum\limits_{\lambda_1\lambda_2} |M(B^\ast\to M_1M_2)|^2
\label{eq:width}
\en
where $\lambda=-\lambda_1+\lambda_2$.
In Table~\ref{tab:nonlepdecay} the branching fractions of all
weak nonleptonic decays considered here are given. As wide accepted,
we used the calculated values of the radiative decay widths as total widths,
i.e. $\Gamma_{\rm tot}(B^\ast)=\Gamma(B^\ast\to B\gamma$). For comparison,
we give the results obtained in Ref.~\cite{Chang:2019xtj}.
\begin{table}[H]
  \caption{Branching fractions of the $B^\ast\to D^\ast V$ decays
($V=\rho,K^\ast,D^\ast,D_s^\ast$)}
\label{tab:nonlepdecay}
\vskip 2mm
\def\arraystretch{1.1}
\centering
\begin{tabular}{|c|c|c|}
\hline
$\qquad$ Decay mode $\qquad$	& $\qquad$ CCQM $\qquad$ &
$\qquad\quad$ CLFQM~\cite{Chang:2019xtj} $\qquad\quad$\\ 
		\hline 
\(B^{*-} \to D^{*0} + K^{*-}\)	&
\(1.02(15) \times 10^{-9} \) &  \(1.10^{+0.01+0.19}_{-0.01-0.17} \times 10^{-11}\)
\\ 
\(B^{*-} \to D^{*0} + \rho^{-}\)	&
\( 1.92(29) \times 10^{-8} \) &  \(2.23^{+0.04+0.39}_{-0.04-0.35} \times 10^{-10}\)
\\
\(B^{*-} \to D^{*0} + D^{*-}\)	&
\( 1.75(26) \times 10^{-9} \) & \(1.44^{+0.11+0.24}_{-0.11-0.22} \times 10^{-11}\)
\\
\(B^{*-} \to D^{*0} + D^{*-}_s\)	&
\( 3.85(58) \times 10^{-8} \) & \(3.71^{+0.18+0.64}_{-0.18-0.57} \times 10^{-10}\)
\\ 
\(\bar{B}^{*0} \to D^{*+} + K^{*-}\) &
\( 4.62(69) \times 10^{-9} \) & \(3.40^{+0.24+0.58}_{-0.23-0.52} \times 10^{-11}\)
\\ 
\(\bar{B}^{*0} \to D^{*+} + \rho^{-}\)	&
\( 8.09(1.21) \times 10^{-8} \) & \(6.85^{+0.26+1.17}_{-0.26-1.05} \times 10^{-10}\)
\\ 
\(\bar{B}^{*0} \to D^{*+} + D^{*-}\) &
\( 5.12(77) \times 10^{-9} \) & \(4.33^{+0.33+0.74}_{-0.32-0.66} \times 10^{-11}\)
\\ 
\(\bar{B}^{*0} \to D^{*+} + D^{*-}_s\) &
\( 1.17(18) \times 10^{-7} \) &  \(1.11^{+0.06+0.19}_{-0.05-0.17} \times 10^{-9}\)
\\ 
\(\bar{B}^{*0}_s \to D^{*+}_s + K^{*-}\) &
\( 5.63(84) \times 10^{-9} \) & \(4.80^{+0.34+0.83}_{-0.32-0.74} \times 10^{-11}\)
\\ 
\(\bar{B}^{*0}_s \to D^{*+}_s + \rho^{-}\) &
\( 5.21(78) \times 10^{-9} \) & \(9.39^{+0.36+1.63}_{-0.35-1.46} \times 10^{-10}\)
\\ 
\(\bar{B}^{*0}_s \to D^{*+}_s + D^{*-}\)	&
\( 6.51(98) \times 10^{-9} \) &  \(6.10^{+0.47+1.03}_{-0.45-0.92} \times 10^{-11}\)
\\ 
\(\bar{B}^{*0}_s \to D^{*+}_s + D^{*-}_s\)	&
\( 1.48(22) \times 10^{-7} \) & \(1.54^{+0.08+0.26}_{-0.07-0.24} \times 10^{-9}\)
\\ 
\hline
\end{tabular}
\end{table}
One can see from Table~\ref{tab:nonlepdecay} that our results are almost
two order larger in magnitude than those from Ref.~\cite{Chang:2019xtj}.
Unfortunately, the authors of \cite{Chang:2019xtj} did not give
the numerical coefficients for the Wilson coefficients $C_1$ and $C_2$,
there are only the combinations of $\alpha_1=C_1+\xi C_2$ and
$\alpha_2=C_2+\xi C_1$ with $\xi=1/N_c$. However, they refer to the
papers by Buras et al.~\cite{Buchalla:1995vs,Buras:1998raa} in which
the notation are accepted as $C_2=1$  and
$C_1=0$ if QCD is neglected.
We are using the Wilson coefficients
from Ref.~\cite{Descotes-Genon:2013vna} (see, Table 3, and references therein)
where  their values have been calculated by using the various corrections
and found that $C_1  = -0.2632$ and $ C_2  = 1.0111$.
If one uses these values for $\alpha_1$ and $\alpha_2$ of
Ref.~\cite{Chang:2019xtj}  then one arrives at
$\alpha_1\approx 0.074$ for $N_c=3$, i.e. strongly
suppressed, whereas the value of  $\alpha_2=0.923$ is of the leading order.
If compare the analytical expressions for the decay amplitudes from
\cite{Chang:2019xtj} (see, Eqs.~(13)-(24)) with our  results given by
Eq.~\ref{eq:ampl-Krho}  and Table~\ref{tab:ampl-charge},
then one finds that in Ref.~\cite{Chang:2019xtj}
the amplitudes with the charged emitted mesons are proportinal to
$\alpha_1$ which is suppressed whereas in our approach those amplitudes
are proportinal to $a_1=C_2+\xi C_1$ which is of leading order.
It could explain two order  difference for  branching fractions obtained
in these two approaches.

\section{Summary}
\label{sec:summary}

The radiative and weak nonleptonic decays of vector $B$-mesons
have been studied within the covariant confined quark model (CCQM) developed
in our previous papers. The matrix elements and decays widths
of the radiative decays $B^\ast\to B\gamma$ were calculated.
The obtained results were compared with those obtained in other approaches.

The nonleptonic decays $B^\ast\to D^\ast V$ which proceed via tree-level quark
diagrams have been carefully analized. It was shown that the analytical
expressions for the amplitudes correspond to the factorization approach.
In the framework of our approach we calculated the leptonic decay constants
and the form factors of the $B^\ast\to D^\ast(V)$ transitions in the entire
physical region of the momentum transfer squared.
Finally, we calculated the two-body decay widths
and compared our results with other models.

\section{Acknowledgements}
\label{sec:acknowledgements}

This work is supported by the JINR grant of young scientists and
specialists No. 22-302-06. Zh.T.’s research has been funded by
the Science Committee of the Ministry of Education and
Science of the Republic of Kazakhstan (Grant No. AP09057862).

\ed